\documentclass[usenatbib]{basi}
\usepackage[T1]{fontenc}
\usepackage[british]{babel}
\usepackage[varg]{txfonts}
\usepackage{natbib}
\conference{The Metrewavelength Sky}

\usepackage{dcolumn}
\begin{document}
\title[First results from CLoGS]{Radio properties of nearby groups of galaxies }
\author[K. Kolokythas et~al.]%
       {K. Kolokythas$^1$\thanks{email: \texttt{kkolok@star.sr.bham.ac.uk}},
       E.~O'Sullivan$^{2}$, S.~Raychaudhury$^{3,1}$, C.~H. Ishwara-Chandra$^{4}$ \& N.~Kantharia$^{4}$\\
       $^1$School of Physics and Astronomy, University of Birmingham, Birmingham, B15 2TT, UK\\
       $^2$Harvard-Smithsonian Center for Astrophysics, 60 Garden Street, Cambridge, MA 02138, USA\\
       $^3$Department of Physics, Presidency University, 86/1 College Street, 700 073 Kolkata, India\\
       $^4$National Centre for Radio Astrophysics,Tata Institute of Fundamental Research, Post Bag 3,\\ Ganeshkhind, 411 007 Pune, India}

\pubyear{2014}
\volume{00}
\pagerange{\pageref{firstpage}--\pageref{lastpage}}

\date{Received --- ; accepted ---}

\maketitle
\label{firstpage}

\begin{abstract}
Much of the evolution of galaxies takes place in groups where feedback has the greatest impact on galaxy formation. By using an optically selected, statistically complete sample of 53 nearby groups (CLoGS), observed in Radio (GMRT) and in X-rays (Chandra \& XMM-Newton), we aim to characterize the radio-AGN population in groups and examine their impact on the intra-group gas and member galaxies. The sensitivity to older electron populations at 240 MHz and the resolution of 610 MHz is the key to identify past and current AGN activity. Here we will present first results from three different galaxy groups analysed so far. We report an age of $\sim$18 Myr for the radio source 3C270 in NGC~4261 implying that the expansion was supersonic over a large fraction of its lifetime. In NGC~1060 we detect a small-scale (20''/7.4 kpc) jet source with the spectral index of $\alpha_{240}^{610}$=0.9 indicating a remnant of an old outburst. Lastly in NGC~5982 the 610 \& 235 MHz observations find a radio point source in the central AGN with no evidence of jets or lobes and diffuse emission from the disks (star formation). 

\end{abstract}


\section{Introduction}\label{s:intro}

Much of the evolution of galaxies takes place in groups where the balance between hot and cold gas becomes significant for the investigation of the hot intra-group medium (IGM)/active galactic nuclei (AGN) connection. The transfer of energy though between AGN and the IGM still remains an unresolved issue. Since most of the present-day galaxies and baryonic matter in the Universe reside in groups (\citealt{Eke2004}, \citealt{Mulchaey2000}) the study of feedback across a wide variety of nearby groups is fundamental, if we are to understand how it has influenced the thermal history of IGM. Such a study requires multi-frequency radio data and high-quality X-ray observations in order to provide insight into the physical mechanisms of the energy injection. 

We define an optically selected, statistically complete sample of 53 nearby groups (Complete Local-Volume Groups Sample project), observed in Radio using GMRT in dual-frequency 610/240 MHz and in X-rays using the Chandra and XMM-Newton satellite observatories, very carefully screened to exclude systems which are uncollapsed (see \citealt{OSullivan2014} for more details of sample selection). Using archival radio data and new observations, we target the characterization of the radio-AGN population in groups of galaxies and the examination of their impact on the intra-group gas and member galaxies.  

The examination of low-frequency radio emission is particularly important in order to identify past as well as current AGN activity, combining the good spatial resolution of the GMRT at 610 MHz and the sensitivity to older electron populations at 240 MHz. Multi-frequency observations also allow the production of spectral index maps. Using a variety of systems which host different types of AGN, the possible mechanisms of AGN feedback heating in groups, the length of the AGN duty cycle and the relationship between environment and AGN jet/lobe properties can be constrained. 

Here results from three different systems that exhibit the apparent diversity of the sample will be presented; We examine a group dominated by a single powerful central source where large-scale jets inflate radio lobes (NGC4261), a group combining weak AGN and star-formation (NGC5985) and a recently confirmed merger group with multiple AGN and star formation detections (NGC1060).

\section{NGC~4261 (3C270)}

\begin{figure}[h]

\centerline{\includegraphics[width=0.52\textwidth]{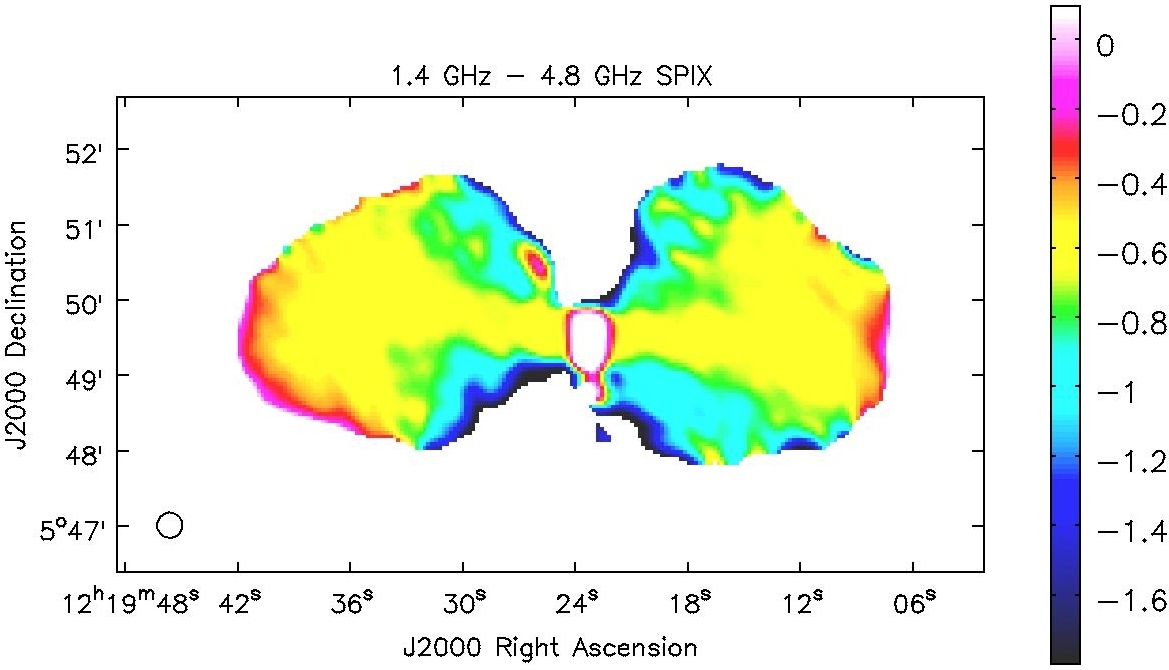}
\includegraphics[width=0.61\textwidth]{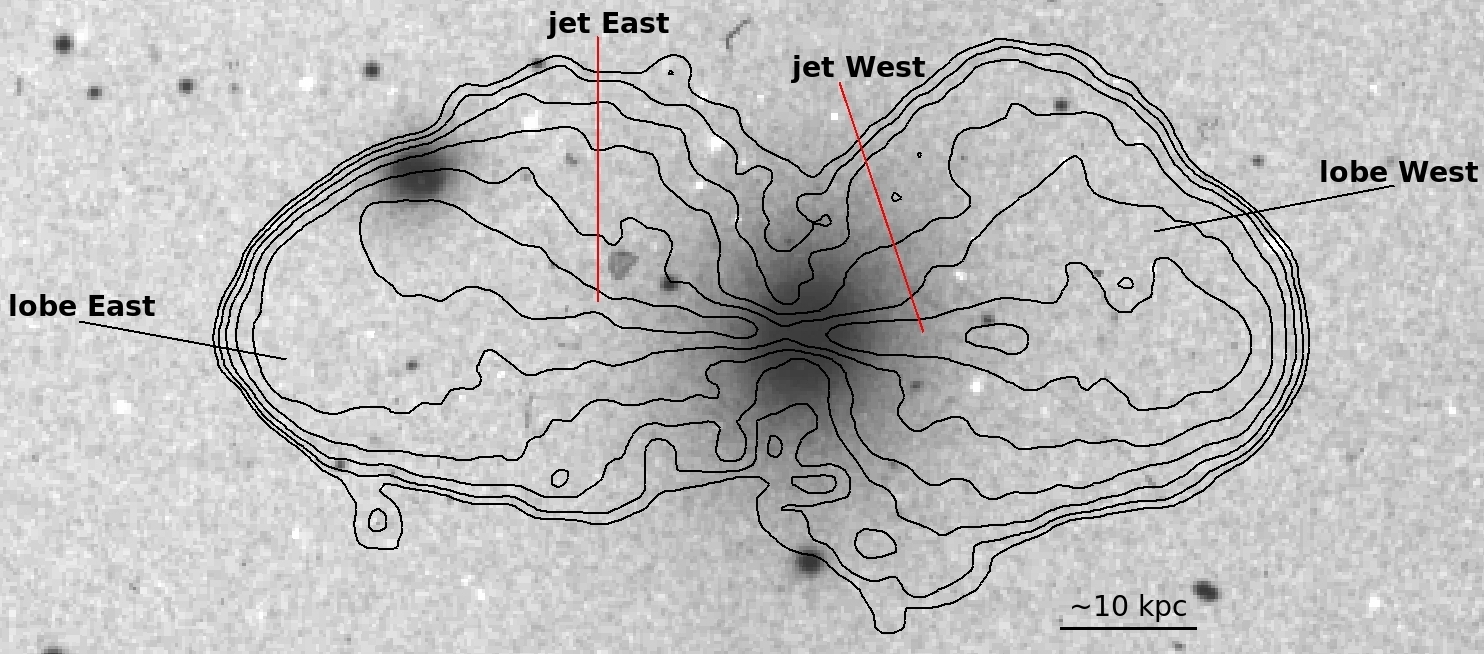}
}
\caption{1.5 GHz-4.8 GHz spectral index distribution over 3C 270 (\textit{left}) produced from images with a restoring beam of 20$^{''} \times 20^{''}$. GMRT 240 MHz (1$\sigma$ = 1.6 mJy beam$^{-1}$) contours of 3C 270, overlaid on the r-band SDSS image (\textit{right}). Contour levels (black) start at 4$\sigma$ and rise by factor of 2.}

\vspace{-0.1cm}
\end{figure}

NGC 4261 hosts 3C 270, an FR-I radio galaxy and the strongest radio source in our sample (19 Jy at 1.4 GHz, \citealt{Kuhr1981}) whose twin jets lie close to plane of the sky \citep{Worrall2010}. We analysed archival VLA and GMRT observations of the source and produced spectral index maps. In Fig. 1 the constant spectral index ($\sim$0.6) of the jets suggests the internal plasma flows are still fast, with the plasma ageing very little over its travel time along the jets. The steeper values at the back of the lobes indicate that the plasma close to the core is actually older than that in the hotspots, suggesting that it was transported out to the tips of the lobes by the jets and has then flowed all the way back in to the galaxy core. 


%


From a point-to-point spectral index distribution analysis (\citealt{Kolokythas2014} in prep.) the age of the outburst was estimated. The radio age that we find from our best fit JP model analysis is $\sim$13 Myr for the west lobe and $\sim$18 Myr for the east one. This is lower than the 75 Myr of the X-ray analysis in \citealt{ewan4261}, implying that the source expansion must have been supersonic over a larger fraction of it's lifetime, with stronger shock heating of the IGM. 



\section{Newly confirmed group: NGC~1060 (LGG~72)}

A new XMM-Newton pointing for the group detects the AGN in several of the galaxies, and reveals a previously unknown extended gaseous halo in the group. We find a characteristic temperature of $\sim$1 keV and luminosity L$_X$ $\sim2.6\times10^{42}$ erg s$^{-1}$, typical of a moderate mass group. 

\begin{figure}[h]
\centerline{
\includegraphics[width=1.0\textwidth]{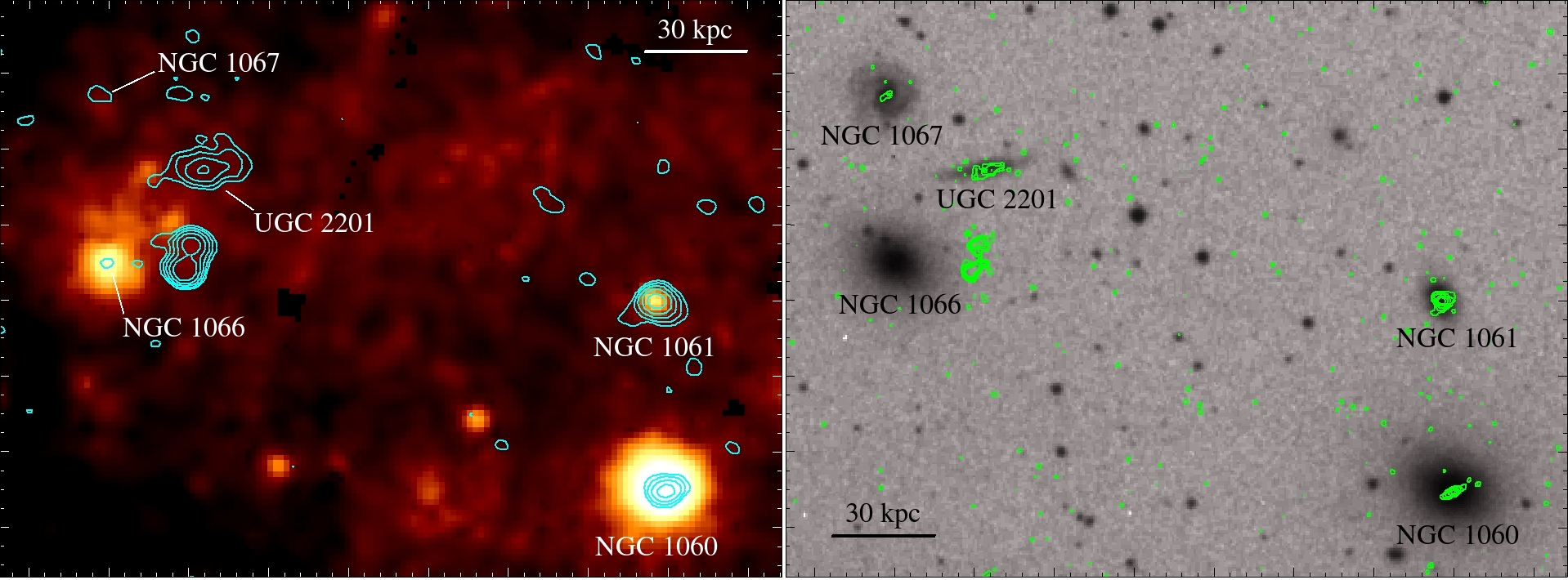}
}
\caption{235 MHz contours (1$\sigma$ = 0.5 mJy beam$^{-1}$) overlaid on the smoothed \textit{XMM-Newton} 0.3-2 keV image (\textit{left}) and 610 MHz contours (1$\sigma$ = 90 $\mu$Jy beam$^{-1}$) overlaid on the r-band SDSS image (\textit{right})}
\end{figure}
However, we find that the IGM is highly disturbed, with separate X-ray peaks centred on the two main ellipticals, NGC 1060 and NGC 1066, and a $\sim$250 kpc arc of hot gas linking the two.
From the GMRT analysis in NGC 1060 we detect a small-scale (20''/7.4 kpc) $\sim$13mJy jet source at 610 MHz ($\sim$30mJy at 240 MHz). The spectral index $\alpha_{240}^{610}=$0.9 indicates a remnant of an old, low power outburst. The system appears to be undergoing a group-group merger, which may have triggered the nuclear activity in NGC 1060. \\

\section{Diversity in Radio: NGC~5982 (Draco Trio)}


\begin{figure}[h]
\centering{
\includegraphics[width=0.95\textwidth]{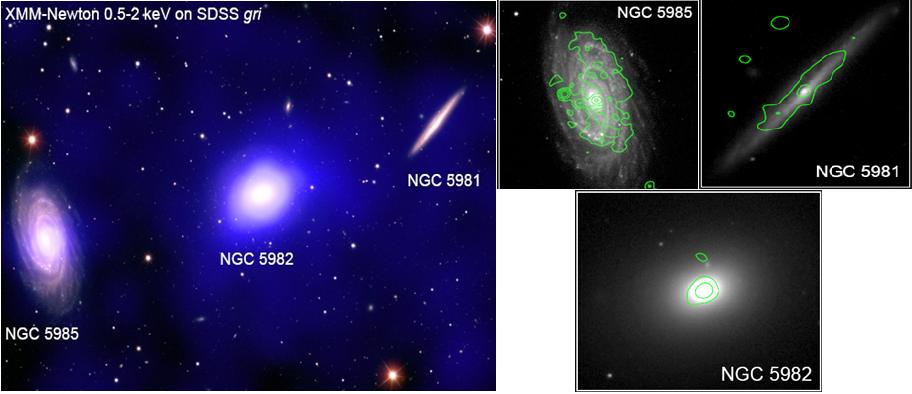}
}
\caption{In blue the \textit{XMM-Newton} 0.5-2 keV on gri SDSS image (\textit{left}) and in green the 610 MHz contours (1$\sigma$ = 90 $\mu$Jy beam$^{-1}$) overlaid on r-band SDSS image (\textit{right})}

\end{figure}

LGG~402 is a relatively poor nearby group (D=44 Mpc) dominated by the elliptical NGC 5982 and two large spiral galaxies, NGC 5981 and NGC 5985. GMRT 610 and 235 MHz observations (Fig. 3) detect weak, point-like AGN in all three galaxies and star-formation in the two disk galaxies. The emission from the two spirals, suggests that the diffuse emission from the disks (star formation) was developed independently from the central source. The central AGN in NGC~5982 is only a radio point source with no evidence of jets or lobes. The spectral index of $\alpha_{240}^{610}=$1 indicates an old, low power AGN. A short XMM pointing reveal a $\sim$0.5 keV IGM centred on the elliptical and detected out to 85 kpc, approximately the projected distance to the neighbouring spirals. The IGM appears relaxed with no obvious substructure suggesting that heating or disruption of a cool core by a merger or AGN outburst is unlikely.




\label{lastpage}
\end{document}